\documentclass[aps,pra,showpacs,superscriptaddress]{revtex4}%
\usepackage{amsmath,amssymb,graphics,epsfig}

\begin{document}

\title{Hardy's proof of nonlocality in the presence of noise}

\author{GianCarlo \surname{Ghirardi}}\email{ghirardi@ts.infn.it}%
\affiliation{Department of Theoretical Physics, University of
  Trieste, Italy}%
\affiliation{Istituto Nazionale di Fisica Nucleare, Sezione di Trieste,
  Italy}%
\affiliation{International Centre for Theoretical Physics ``Abdus Salam,''
  Trieste, Italy}%

\author{Luca \surname{Marinatto}}\email{marinatto@ts.infn.it}%
\affiliation{Department of Theoretical Physics, University of
  Trieste, Italy}%
\affiliation{Istituto Nazionale di Fisica Nucleare, Sezione di Trieste,
  Italy}%

\date{\today}

\begin{abstract}
We extend the validity of Hardy's nonlocality without inequalities proof
 to cover the case of special one-parameter classes of nonpure statistical operators.
These mixed states are obtained by mixing the Hardy states with
 a completely chaotic noise or with a colored noise and they represent a realistic description of
 imperfect preparation processes of (pure) Hardy states in nonlocality experiments.
Within such a framework we are able to exhibit a precise range of values of the parameter measuring
 the noise affecting the non-optimal preparation of an arbitrary Hardy state, for
 which it is still possible to put into evidence genuine nonlocal effects.
Equivalently, our work exhibits particular classes of bipartite mixed states whose
 constituents do not admit any local and deterministic hidden variable model reproducing
 the quantum mechanical predictions.
\end{abstract}

\pacs{03.65.Ud}

\maketitle

\section{Introduction}

Nonlocality without inequalities arguments, like the celebrated Greenberger-Horne-Zeilinger (GHZ)~\cite{ghz}
 and Hardy's~\cite{hardy} ones, may provide evidence of the genuine nonlocal features of
 certain quantum states without resorting to the consideration of Bell-like inequalities~\cite{bell,chsh}.
The experimental verification of the implications of such arguments does not require one to collect data of
 several correlation functions but it simply demands one to test the occurrence of certain joint
 measurement outcomes.
More precisely, in the case of the Greenberger-Horne-Zeilinger argument one has to verify the existence of
 perfect (anti)correlations between the outcomes of dichotomic measurements performed on to each of a group
 of three spacelike separated particles~\cite{bouwmeester,pan}, while in the case of
 Hardy's proof one has to put into evidence the occurrence of a single joint event~\cite{demartini,hardy2}.
Such experiments usually involve groups of polarization-entangled photons and the most
 reliable source of (multipartite) entanglement is represented by parametric down
 conversion processes~\cite{kwiat}.
However, in practice, the pure states exhibiting nonlocal effects one wishes to prepare in a laboratory
 in this way (the tripartite GHZ state and the Hardy states, which are bipartite state vectors whose Schmidt
 decomposition involves at least two different weights) are unavoidably subjected to different kinds of noise.
The aim of this paper is to show how it is possible to generalize the original
 Hardy's proof~\cite{hardy} by following techniques which are similar to those we used in Refs.~\cite{gm,gm2}
 and to apply it to specific one-parameter classes of statistical operators, representing mixtures
 of Hardy states with a completely chaotic noise (also known as a white noise) or with a colored noise.
While the first kind of noise the one which is usually considered in the literature for nonlocality
 experiments, the second kind
 has been recently pointed out to be the best (and more realistic) choice for describing what really
 happens in type II spontaneous parametric down conversion processes~\cite{cabe}.

Therefore, in this paper, we will determine the range of values of the parameter which measures
 the amount of noise affecting the nonoptimal preparation of a definite (pure) Hardy state, for which
 it is still possible that the resulting mixed state exhibits genuine and testable nonlocal effects.
Equivalently, our work exhibits particular classes of bipartite mixed states whose
 constituents do not admit any local and deterministic hidden variable model which may consistently
 be in agreement with the quantum mechanical predictions, and which are, as a consequence, nonseparable
 (that is, they cannot be expressed as a convex sum of product states).
As already remarked on, the techniques we use are similar to those presented in Ref.~\cite{gm2} but some
 results are improved because in this paper we exploit the explicit form of the noise corrupting a Hardy state,
 while in Ref.~\cite{gm2} the form of the mixed states we were considering was left unspecified for the sake
 of generality.


\section{Hidden variable models}

In ordinary quantum mechanics, given the state vector associated to a physical system,
 the outcomes of (single and joint) measurement processes of arbitrary observables may be only statistically
 predicted and, apart from particular situations, such outcomes are not certain.
However, there still exists the (logical) possibility that the quantum states
 do not represent the maximal possible knowledge one can have about a quantum system.
In such a situation, supplementary variables might very well predetermine the outcomes
 of any quantum measurement and the stochastic aspects of quantum mechanics would arise since
 what one can know is only the probabilistic distribution of these additional
 variables ---for this reason they are referred to as hidden variables.

Accordingly, the hypothetical theory (called stochastic hidden variable model) which completes quantum mechanics
 consists of (i) a set $\Lambda$, whose elements $\lambda$ are the hidden variables; (ii) a normalized
 probability distribution $\rho$ defined on $\Lambda$; (iii) a set of
 probability distributions $P_{\lambda}(A_{i}\!=\!a, B_{j}\!=\!b,\dots, Z_{k}\!=\!z)$
 for the outcomes of single and joint measurements of any conceivable
 set of observables $\left\{A_{i},B_{j},\dots,Z_{k} \right\}$,
 where each index of the set $\left\{ i,j,\dots,k\right\}$ refers to a single particle or to
 a group of all the particles, such that
\begin{equation}
\label{eq0.1} P_{\sigma}(A_{i}=a,B_{j}=b,\dots,Z_{k}=z) = \int_{\Lambda}\,d\lambda\,
   \rho(\lambda) P_{\lambda}(A_{i}=a,B_{j}=b,\dots,Z_{k}=z).
\end{equation}
The quantity at the left hand side of Eq.~(\ref{eq0.1}) is the quantum mechanical
 (joint) probability distribution for the set of outcomes $\left\{ a,b, \dots, z\right\}$ of the
 considered measurements when the system is in the state $\sigma$.
A deterministic hidden variable model, which is also known as a realistic model, is a
 particular instance of a hidden variable model where all probabilities $P_{\lambda}$ take
 the values $0$ or $1$ only.
In this case, the measurement outcomes of arbitrary observables are predetermined.

A hidden variable model is called local~\cite{bell2} if the following factorizability condition
 holds for any conceivable joint probability distribution $P_{\lambda}(A_{i}=a,B_{j}=b, \ldots, Z_{k}=z)$
 and for any value of the hidden variable $\lambda\in \Lambda$
\begin{equation}
 \label{eq0.2}
 P_{\lambda}(A_{i}=a,B_{j}=b,\ldots,Z_{k}=z)= P_{\lambda}(A_{i}=a)P_{\lambda}(B_{j}=b)\ldots
 P_{\lambda}(Z_{k}=z)
 \end{equation}
in all cases in which the measurement processes for the observables $A_{i},B_{j},\ldots, Z_{k}$
 occur at spacelike separated locations. The locality condition imposes that no causal influence
 can exist between spacelike separated events.
Since deterministic and stochastic hidden variable models are totally equivalent when the locality
 condition is imposed~\cite{fine}, and since in what follows we will focus only on realistic models,
 all the probabilities $P_{\lambda}$ will consequently be assumed to take the values $0$ and $1$ only.


\section{Hardy's proof in the presence of noise}

A Hardy state $\vert \psi \rangle$ belonging to ${\mathbb{C}}^{2}\otimes {\mathbb{C}}^{2}$ is any entangled
 state whose Schmidt decomposition with respect to an orthonormal basis $\left\{ \vert 0\rangle,\vert
 1\rangle\right\}$, involves two (strictly positive) different weights $p_{1}\neq p_{2}$ so that
\begin{equation}
\label{eq1.1}
 \vert \psi \rangle = p_{1}\vert 0\rangle \otimes\vert 0\rangle +p_{2}\vert 1\rangle \otimes\vert 1
 \rangle, \qquad p_{1}^{2}+p_{2}^{2}=1\:.
\end{equation}
Thus, the Hardy states of Eq.~(\ref{eq1.1}) are simply all entangled but not maximally entangled two-qubit
 states. By following Ref.~\cite{gm} we define two orthonormal bases, depending on the
 Schmidt coefficients $p_{1}$ and $p_{2}$ of Eq.~(\ref{eq1.1}),
 $\left\{ \vert x_{+} \rangle, \vert x_{-}\rangle \right\}$
 and $\left\{ \vert y_{+}\rangle, \vert y_{-}\rangle \right\}$ spanning ${\mathbb{C}}^{2}$, as follows:
\begin{equation}
\label{eq1.31}
\begin{bmatrix} \vert x_{+} \rangle \\  \vert x_{-} \rangle \end{bmatrix}
=\frac{1}{\sqrt{p_{1}+p_{2}}}
\begin{bmatrix}
\sqrt{p_{2}} & -i\sqrt{p_{1}} \\
-i \sqrt{p_{1}} & \sqrt{p_{2}}
\end{bmatrix}
\begin{bmatrix} \vert 0 \rangle \\  \vert 1 \rangle \end{bmatrix}
\end{equation}
\begin{equation}
\label{eq1.32}
\begin{bmatrix} \vert y_{+} \rangle \\  \vert y_{-} \rangle \end{bmatrix}
= \frac{1}{\sqrt{(p_{1}^{2}+p_{2}^{2}-p_{1}p_{2})(p_1+p_2)}}
\begin{bmatrix}
-ip_2\sqrt{p_2} & p_1\sqrt{p_{1}} \\
p_1\sqrt{p_{1}} & -ip_{2}\sqrt{p_2}
\end{bmatrix}
\begin{bmatrix} \vert 0 \rangle \\  \vert 1 \rangle \end{bmatrix}\:.
\end{equation}
Then, we denote as $X_{i}$ and $Y_{i}$ (where $i=1,2$ is the particle index) the observables
 whose eigenstates associated to the eigenvalues $+1$ and $-1$ are the
 vectors $\left\{ \vert x_{+} \rangle, \vert x_{-}\rangle \right\}$
 and $\left\{ \vert y_{+}\rangle, \vert y_{-}\rangle \right\}$ of Eqs.~(\ref{eq1.31}) and~(\ref{eq1.32}), respectively.

State vectors $\vert \psi \rangle$ of the kind of Eq.~(\ref{eq1.1}) are known to exhibit nonlocal
 features~\cite{hardy} because certain joint probability distributions involving the observables
 $X_{i}$ and $Y_{i}$ turn out to be locally inexplicable.
The aim of this paper is to prove that similar nonlocal effects arise also when considering particular classes
 of mixed states generated from the Hardy states of Eq.~(\ref{eq1.1}).
To start with, let us consider the mixed statistical operator obtained by taking the convex sum of a pure
 Hardy state of Eq.~(\ref{eq1.1}) with a completely chaotic noise (also called white noise) as
\begin{equation}
\label{eq1.4}
 \sigma = p\vert \psi \rangle \langle \psi \vert + \frac{1-p}{4} I_{2}\otimes I_{2}\:,
\end{equation}
where $I_{2}$ is the identity operator in ${\mathbb{C}}^{2}$ and $p\in [0,1]$ is the parameter
 which measures the amount of noise affecting the purity of the Hardy state $\vert \psi\rangle$.
Such a statistical operator represents the state which usually results from an imperfect preparation
 of the (pure) Hardy state $\vert \psi\rangle$ in the practical realizations of nonlocality experiments.
In fact, it is customary to mimic the noise which affects a nonperfectly
 controllable preparation of pure entangled states, like those of polarization-entangled photons produced
 by parametric down-conversion processes, as the output state of Eq.~(\ref{eq1.4})~\cite{white}.

In order to exhibit a proof of nonlocality without inequalities for the state $\sigma$ by following the Hardy's
 argument, we have to consider appropriate quantum mechanical probability distributions and to show that they
 cannot be reproduced by local realistic hidden variable models.
To this end let us briefly review the line of reasoning of Ref.~\cite{gm2} and consider the observables $X_{i}$
and $Y_{i}$, defined above, and we take into account the
 following joint-probability distributions when the state of the system is that of Eq.~(\ref{eq1.4}):
\begin{eqnarray}
 \label{eq1.51}
 P_{\sigma}(X_{1}=+1, X_{2}=+1) &=& \frac{1-p}{4}\equiv \varepsilon, \\
 \label{eq1.52}
 P_{\sigma}(Y_{1}=+1, X_{2}=-1) &=& \frac{1-p}{4}\equiv \varepsilon, \\
 \label{eq1.53}
 P_{\sigma}(X_{1}=-1, Y_{2}=+1) &=& \frac{1-p}{4}\equiv \varepsilon, \\
 \label{eq1.54}
 P_{\sigma}(Y_{1}=+1, Y_{2}=+1) &=& p
  \frac{p_{1}^{2}p_{2}^{2}(p_{1}-p_{2})^2}{(1-p_{1}p_{2})^{2}} +\frac{1-p}{4} \equiv a +\varepsilon\:,
\end{eqnarray}
where the quantity $\frac{p_{1}^{2}p_{2}^{2}(p_{1}-p_{2})^2}{(1-p_{1}p_{2})^{2}}$ does not vanish (in fact, by
 hypothesis, $p_{1}$ and $p_{2}$ are strictly positive and different from each other).
This quantity represents the crucial joint probability which allows one to establish the nonlocal features of
any
 Hardy state in the original formulation of the argument of nonlocality~\cite{hardy,gm}.

Now we are going to determine which are the allowed values for the parameter $p$ if a local and realistic model
 exists which is able to  reproduce the probabilities of Eqs.~(\ref{eq1.51})-(\ref{eq1.54}).
To this purpose, suppose that a local and realistic description (as defined in the previous section)
 exists for the state $\sigma$ and consider what happens, for example, with the probability
 distribution $P_{\sigma}(X_{1}=+1,X_{2}=+1)$.
According to the definition of a local hidden variable model, we have
\begin{eqnarray}
\label{eq1.6}
  P_{\sigma}(X_{1}=+1,X_{2}=+1)
  & = & \int_{\Lambda}\,d\lambda\, \rho(\lambda) P_{\lambda}(X_{1}=+1)P_{\lambda}(X_{2}=+1).
\end{eqnarray}
Since we are dealing with a deterministic model, where the (single-particle) probabilities $P_{\lambda}$
 possess values $0$ or $1$ only, it is useful to define the following subsets $A,B,C,$ and $D$ of $\Lambda$ as
\begin{eqnarray}
 \label{eq1.71}
 A &= &\left\{ \:\lambda\in \Lambda \:\vert \:P_{\lambda}(X_{1}=+1)=1\right\},\\
\label{eq1.72}
 B &= &\left\{ \:\lambda\in \Lambda \:\vert \:P_{\lambda}(X_{2}=+1)=1\right\},\\
\label{eq1.73}
 C &= &\left\{ \:\lambda\in \Lambda \:\vert \:P_{\lambda}(Y_{1}=+1)=1\right\},\\
\label{eq1.74}
 D &= &\left\{ \:\lambda\in \Lambda \:\vert \:P_{\lambda}(Y_{2}=+1)=1\right\}.
\end{eqnarray}
If we denote as $\mu(\Sigma)$ the measure of any subset $\Sigma$ of $\Lambda$
 with respect to the weight function $\rho(\lambda)$, i.e., $\mu(\Sigma)=\int_{\Sigma}\,d\lambda\,
 \rho(\lambda)$, the probability distributions of Eqs.~(\ref{eq1.51})-(\ref{eq1.54}) turn out to be equivalent to
\begin{eqnarray}
 \label{eq1.91}
 \mu[A \cap B] &= & \varepsilon, \\
 \label{eq1.92}
 \mu[C] - \mu[ B\cap C] & = & \varepsilon,\\
\label{eq1.93}
 \mu[D]- \mu[A\cap D] & = & \varepsilon,\\
\label{eq1.94}
 \mu[ C\cap D] & = & a+\varepsilon.
\end{eqnarray}
If we follow the set-theoretic manipulations presented in Ref.~\cite{gm2}, starting from
 Eqs.~(\ref{eq1.91})-(\ref{eq1.94}), we end up with an inequality constraining the values of $\varepsilon$ and
 $a$, as long as a local realistic model for $\sigma$ is supposed to exist, written
\begin{equation}
\label{eq1.95}
 2\varepsilon - a \geq 0.
\end{equation}
This relation, when expressed in terms of the parameters $p$, $p_{1}$ and $p_{2}$, takes the
 following form
\begin{equation}
\label{eq2.3} 0 \leq  p \leq \frac{1}{1+ 2\frac{p_{1}^{2}p_{2}^{2}(p_{1}-p_{2})^2}{(1-p_{1}p_{2})^{2}}}\:.
\end{equation}
To summarize, we have proven the following theorem:\\

{\bf Theorem I}. Consider the normalized entangled vector $\vert \psi \rangle = p_{1}\vert 0\rangle\vert
0\rangle +
 p_{2}\vert 1\rangle \vert 1 \rangle$ belonging to ${\mathbb{C}}^{2}\otimes {\mathbb{C}}^{2}$,
  with different (strictly positive) weights $p_{1}\neq p_{2}$, and the one-parameter
 class of mixed statistical operators $\sigma = p\vert \psi\rangle \langle \psi
 \vert + \frac{1-p}{4}I_{2}\otimes I_{2}$, where $p\in [0,1]$.
If there exists a local and deterministic hidden variable model for $\sigma$ then
 $ p \in [0, \frac{1}{1+ 2\frac{p_{1}^{2}p_{2}^{2}(p_{1}-p_{2})^2}{(1-p_{1}p_{2})^{2}}}]$.\\

Thus, we have succeeded in exhibiting a necessary condition for the existence of a local realistic model for
 the one-parameter class of operators $\sigma$ of Eq.~(\ref{eq1.4}).
The usefulness of this result is twofold, both from the theoretical and from the experimental point of view, as
can be immediately deduced by reversing the assertion of the previous theorem.

In fact, first, for any $p \in (\frac{1}{1+ 2\frac{p_{1}^{2}p_{2}^{2}(p_{1}-p_{2})^2}{(1-p_{1}p_{2})^{2}}},1]$
 the associated mixed states cannot be locally described in terms of a deterministic hidden variable model
 where all the measurement outcomes are predetermined. As a consequence, such states are proven to be
 not separable, that is, they cannot be decomposed as a convex sum of product states (indeed, if they were separable
 a local realistic model for them would actually exist~\cite{werner}).

Second, for the same values of $p$, genuine nonlocal effects can be successfully revealed by
 experiments where joint measurements are performed by spacelike separated observers,
 despite the presence of a completely chaotic noise corrupting a pure Hardy state.
Stated equivalently, we have been able to obtain a (not necessarily optimal) bound for the maximum amount
 of white noise affecting any state $\vert \psi \rangle$ of the type of Eq.~(\ref{eq1.1}),
 in the presence of which it is still possible to exhibit a Hardy's proof of nonlocality.

To give a numerical example let us consider a (pure) Hardy state $\vert \psi \rangle$ of Eq.~(\ref{eq1.1})
 for which $p_{1}p_{2}=(3-\sqrt{5})/2$.
In this case the probability
 $P_{\psi}(Y_{1}=+1,Y_{2}=+1)=\frac{p_{1}^{2}p_{2}^{2}(p_{1}-p_{2})^2}{(1-p_{1}p_{2})^{2}}$
 attains its maximum value,  approximately equal to $0.09$. As a consequence, in this situation
 one obtains the maximal violation of the locality condition allowed by the original Hardy's proof~\cite{hardy}.
By using the result of Eq.~(\ref{eq2.3}) and the above value for $p_{1}p_{2}$, we can conclude
 that, as long as the parameter $p$ belongs to the interval $p\in (0.85,1)$ ---the inferior value of this
 interval being a two-digit approximation of the exact value one can deduce from Eq.~(\ref{eq2.3})---
 one can still put experimentally into evidence nonlocal effects despite the presence of a white
 noise affecting the preparation of the pure Hardy state $\vert \psi\rangle$.

Let us pass now to analyze what happens if we replace the white noise considered so far
 with another kind of noise which has been recently suggested~\cite{cabe} to be a more realistic description for
 a noise affecting the preparation of entangled states which are generated through parametric-down
 conversion processes.
It is referred to as a colored noise and the one-parameter class of mixed states we will now consider is
\begin{equation}
\label{eq3.1}
 \sigma = p\vert \psi \rangle \langle \psi \vert + \frac{1-p}{2} (\:\vert 0\rangle \langle 0 \vert
 \otimes \vert 0\rangle \langle 0 \vert + \vert 1\rangle \langle 1\vert
 \otimes \vert 1\rangle \langle 1 \vert\:)\,
\end{equation}
where, once again, the parameter $p$ measures the degree of purity of the state and it belongs to
 the interval $[0,1]$.
Given the states of Eq.~(\ref{eq3.1}), we can calculate the (modified) joint probabilities for the set of
 observables $X_{i}$ and $Y_{i}$  obtaining:
\begin{eqnarray}
 \label{eq3.51}
 P_{\sigma}(X_{1}=+1, X_{2}=+1) &=& \frac{1-p}{2(p_{1}+p_{2})^{2}}\equiv \varepsilon_{1},\\
 \label{eq3.52}
 P_{\sigma}(Y_{1}=+1, X_{2}=-1) &=& \frac{(1-p)p_{1}p_{2}}{2(p_{1}+p_{2})^{2}(1-p_{1}p_{2})}\equiv \varepsilon_{2},\\
 \label{eq3.53}
 P_{\sigma}(X_{1}=-1, Y_{2}=+1) &=& \frac{(1-p)p_{1}p_{2}}{2(p_{1}+p_{2})^{2}(1-p_{1}p_{2})}\equiv \varepsilon_{2},\\
 \label{eq3.54}
 P_{\sigma}(Y_{1}=+1, Y_{2}=+1) &=&
  \frac{1-3p_{1}^{2}p_{2}^{2}+p(-8p_{1}^{4}p_{2}^{4}+5p_{1}^{2}p_{2}
  ^{2} -1)}{2(p_{1}+p_{2})^{2}(1-p_{1}p_{2})^{2}}
  \equiv \varepsilon_{3}.
\end{eqnarray}
If we suppose once again that there exists a local and deterministic hidden variable model able to reproduce the
 probability distributions of Eqs.~(\ref{eq3.51})-(\ref{eq3.54}), and if we define the subsets $A,B,C$, and $D$
 as before and follow the reasonings presented in Ref.~\cite{gm2}, we end up with an
 inequality constraining the values of $\varepsilon_{1},\varepsilon_{2}$, and $\varepsilon_{3}$,
\begin{equation}
\label{eq3.6}
 \varepsilon_{1} +2\varepsilon_{2} -\varepsilon_{3}\geq 0.
\end{equation}
As a consequence, by using the definition of $\varepsilon_{i}$ as given in Eqs.~(\ref{eq3.51})-(\ref{eq3.54}),
 we obtain a relation, equivalent to that of Eq.~(\ref{eq3.6}), in terms of the parameters $p_{1},p_{2}$, and
 $p$,
\begin{equation}
\label{eq3.61} 0\leq  p \leq \frac{1}{2(1- 2p_{1}^{2}p_{2}^{2})}
\end{equation}
and the following theorem holds:\\

{\bf Theorem II}. Consider the normalized entangled vector $\vert \psi \rangle = p_{1}\vert 0\rangle\vert
0\rangle +
 p_{2}\vert 1\rangle \vert 1 \rangle$ belonging to ${\mathbb{C}}^{2}\otimes {\mathbb{C}}^{2}$,
  with different (strictly positive) weights $p_{1}\neq p_{2}$, and the one-parameter
  class of  mixed statistical operators $\sigma = p\vert \psi\rangle \langle \psi
 \vert +\frac{1-p}{2} (\:\vert 0\rangle \langle 0 \vert \otimes\vert 0\rangle \langle 0 \vert +
 \vert 1\rangle \langle 1\vert \otimes\vert 1\rangle \langle 1 \vert\:) $ where $p\in [0,1]$.
If there exists a local and deterministic hidden variable model for $\sigma$ then
 $ p \in [0, \frac{1}{2(1- 2p_{1}^{2}p_{2}^{2})}]$.

Stated equivalently, whenever $p\in (\frac{1}{2(1- 2p_{1}^{2}p_{2}^{2})},1]$, no local realistic model can exist
 for the corresponding mixed state, describing a Hardy state
 corrupted by a colored noise. As a consequence, such states cannot be separable states.
From the experimental point of view, Hardy states mixed with a colored noise are more useful, with respect
 to those corrupted by a white noise, for what concerns the possibility of highlighting nonlocal
 effects.
In fact, since the following inequalities
\begin{equation}
\label{eq3.7} 0< \frac{1}{2(1- 2p_{1}^{2}p_{2}^{2})} < \frac{1}{1+
2\frac{p_{1}^{2}p_{2}^{2}(p_{1}-p_{2})^2}{(1-p_{1}p_{2})^{2}}} <1
\end{equation}
hold for all values of $p_{1}\neq p_{2}$, such that $p_{1}^{2}+p_{2}^{2}=1$,  when one
 deals with a colored  rather than with a white noise one obtains a larger
 interval of values of $p$, for which nonlocal effects can be experimentally put into evidence.

To give a numerical example, let us consider again the (pure) Hardy state which implies the maximal violation of
 the locality condition in Hardy's proof: since for it $p_{1}p_{2}=(3-\sqrt{5})/2$, one
 obtains by theorem II that for any $p\in (0.70,1]$ it is possible to put into
 evidence nonlocal effects and this is to be compared with the analogous result $p\in
 (0.85,1]$ we have  for the case of a white noise.

Before passing to analyze what happens in Hilbert spaces of greater dimensionality, let us address the issue of
 how powerful the Hardy's criterion is
 in excluding the existence of local realistic models, for the considered mixed states of Eqs.~(\ref{eq1.4})
 and~(\ref{eq3.1}), with respect to other existing criteria.
In fact, besides Hardy's proof, the usual way to reject local hidden variable models for a given quantum state
 consists in exhibiting a Bell inequality which is violated by an appropriate choice of single-particle
 observables.
For this purpose, and in the restricted scenario of a ${\mathbb{C}}^{2}\otimes
 {\mathbb{C}}^{2}$ Hilbert space, a necessary condition has been exhibited in Ref.~\cite{horo} in order
 that a given mixed state satisfies all conceivable CHSH inequalities~\cite{chsh}.
Such a condition, when applied to the states $\sigma$ of Eq.~(\ref{eq1.4}), affirms that $p\in [0,
 \frac{1}{\sqrt{1+4p_{1}^{2}p_{2}^{2}}}]$ if $\sigma$ does not violate any Bell-like inequality.
Equivalently, one can conclude that whenever
\begin{equation}
\label{eq3.71} p\in (\frac{1}{\sqrt{1+4p_{1}^{2}p_{2}^{2}}},1]
\end{equation}
there exists at least one CHSH inequality which is violated, and, consequently,
 that $\sigma$ cannot be described by any local realistic model.
Unfortunately, since the following inequality
\begin{equation}
\label{eq2.4}
 \frac{1}{\sqrt{1+4p_{1}^{2}p_{2}^{2}}} <
 \frac{1}{1+ 2\frac{p_{1}^{2}p_{2}^{2}(p_{1}-p_{2})^2}{(1-p_{1}p_{2})^{2}}}<1,
 \hspace{1cm}
\end{equation}
holds for any (strictly positive) $p_{1}\neq p_{2}$, such that $p_{1}^{2}+p_{2}^{2}=1$,
 it turns out that the criterion
 of Ref.~\cite{horo} is more powerful than our generalized version of Hardy's proof to deny
 the existence of local realistic models for the considered class of mixed states of Eq.~(\ref{eq1.4}).
A similar conclusion can be reached when considering the states of Eq.~(\ref{eq3.1}). In fact,
 in this case, the criterion of Ref.~\cite{horo} tells that for any $p\in (0,1]$ there exists
 a violated CHSH inequality and, as a consequence, local realistic models cannot exist for any value
 of $p\in (0,1]$, while our method individuates nonlocal states only for values of $p$ belonging to
 the interval $(\frac{1}{2(1- 2p_{1}^{2}p_{2}^{2})},1]$.

Thus from a theoretical point of view, our argument of nonlocality is weaker than the one exhibited in
Ref.~\cite{horo} if consideration is given to the restricted one-parameter class of mixed states of
 Eqs.~(\ref{eq1.4}) and~(\ref{eq3.1}).
On the contrary, from a practical point of view, the result obtained when considering white
 noise is of interest since, to identify and to test experimentally the crucial CHSH inequality which is violated
 by those states $\sigma$ such that Eq.~(\ref{eq3.71}) holds, might turn out to be much more difficult than to
 perform the tests which are the crucial ones from our perspective.
Moreover, better results will be achieved when we will soon consider classes of mixed
 states in ${\mathbb{C}}^{d_{1}}\otimes {\mathbb{C}}^{d_{2}}$, where $d_{1}$ and $d_{2}$ can be any positive integer
 greater than or equal to two, a situation in which no criterion of the type of Ref.~\cite{horo} is currently
 known.

To this end, let us now consider an entangled state $\vert \phi \rangle$ whose Schmidt decomposition
 in terms of appropriate orthonormal sets of states
 $\left\{ \vert k \rangle \right\}$ belonging to ${\mathbb{C}}^{d_{1}}$
 and $\left\{ \vert j \rangle \right\}$ belonging to ${\mathbb{C}}^{d_{2}}$,
 respectively, involves at least two (strictly positive) different weights --which we suppose for simplicity
 to be $p_{1}$ and $p_{2}$:
\begin{equation}
\label{eq4.1}
 \vert \phi \rangle = p_{1}\vert 0\rangle \otimes\vert 0\rangle +p_{2}\vert 1\rangle \otimes\vert 1
 \rangle +\sum_{i \geq 3}^{ \leq min(d_{1},d_{2})} p_{i}\vert i-1\rangle \otimes\vert i-1\rangle,
 \:\:\:\:\:\:\sum_{i}p_{i}^{2}=1.
\end{equation}
States like these will again be referred to as Hardy states.
 Subsequently, we define within the two-dimensional manifold spanned by $\left\{\vert 0\rangle,
 \vert 1\rangle\right\}$ the vectors $\left\{ \vert x_{+} \rangle, \vert x_{-}\rangle \right\}$
 and $\left\{ \vert y_{+}\rangle, \vert y_{-}\rangle \right\}$ as in Eqs.~(\ref{eq1.31})-(\ref{eq1.32}).
The observables $X_{j}$ and $Y_{j}$ (where $j=1,2$) are then defined as before, with the further
 condition that they possess the degenerate eigenvalue $0$ (in addition to $+1$ and $-1$) whose eigenmanifold
 is that spanned by the vectors $\left\{\vert i-1\rangle\right\}$ for any $i \geq 3$.

Once again, in order to generalize the results we obtained in the case of bipartite mixed states of two
 spin-$1/2$ particles, let us consider the one-parameter class of mixed statistical
 operators $\sigma$, obtained by taking a convex mixture of a Hardy state $\vert \phi\rangle$ of Eq.~(\ref{eq4.1})
 and a completely chaotic noise as follows:
\begin{equation}
\label{eq4.2}
 \sigma = p\vert \phi \rangle \langle \phi \vert + \frac{1-p}{d_{1}d_{2}} I_{d_{1}}\otimes I_{d_{2}}\:,
\end{equation}
where $I_{d_1}$ and $I_{d_2}$ are the identity operator in ${\mathbb{C}}^{d_1}$ and ${\mathbb{C}}^{d_2}$,
 respectively.
In order to determine a range of values of the parameter $p$ such that one can prove that
 a local and deterministic hidden variable model for the corresponding mixed states cannot exist, we need
 to take into account appropriate correlation functions.
To this end, given the state of Eq.~(\ref{eq4.2}), we will consider the following joint probability
 distributions for the measurement outcomes of the observables $X_{i}$ and $Y_{j}$
\begin{eqnarray}
 \label{eq4.51}
 P_{\sigma}(X_{1}=+1, X_{2}=+1) &=& (1-p)/d_{1}d_{2}\equiv \varepsilon,\\
 \label{eq4.52}
 P_{\sigma}(Y_{1}=+1, X_{2}=-1) &=& (1-p)/d_{1}d_{2}\equiv \varepsilon,\\
 \label{eq4.53}
 P_{\sigma}(X_{1}=-1, Y_{2}=+1) &=& (1-p)/d_{1}d_{2}\equiv \varepsilon,\\
 \label{eq4.54}
 P_{\sigma}(Y_{1}=+1, X_{2}=0) &=& (1-p)/d_{1}d_{2}\equiv \varepsilon,\\
 \label{eq4.55}
 P_{\sigma}(X_{1}=0, Y_{2}=+1) &=& (1-p)/d_{1}d_{2}\equiv \varepsilon,\\
 \label{eq4.56}
 P_{\sigma}(Y_{1}=+1, Y_{2}=+1) &=& p
  \frac{p_{1}^{2}p_{2}^{2}(p_{1}-p_{2})^2}{(p_{1}^{2}+p_{2}^{2}-p_{1}p_{2})^{2}} +\frac{1-p}{d_{1}d_{2}} \equiv
   a+ \varepsilon.
\end{eqnarray}
Let us now suppose that a local realistic model may account for such joint probability
 distributions, define the sets $A,B,C$, and $D$ as in Eqs.~(\ref{eq1.71})-(\ref{eq1.74})
 and let us see what happens with, e.g., Eq.~(\ref{eq4.52}):
\begin{eqnarray}
\label{eq4.6}
  P_{\sigma}(Y_{1}=+1,X_{2}=-1)
  & = & \int_{\Lambda}\,d\lambda\, \rho(\lambda) P_{\lambda}(Y_{1}=+1)P_{\lambda}(X_{2}=-1)\\
 & = & \int_{\Lambda}\,d\lambda\, \rho(\lambda) P_{\lambda}(Y_{1}=+1)[1-P_{\lambda}(X_{2}=1)+
 P_{\lambda}(X_{2}=0)]\\
 &=& \mu[C] -\mu[B\cap C] - \varepsilon.
\end{eqnarray}
The second equality follows since $P_{\lambda}(X_{2}=-1) +P_{\lambda}(X_{2}=0)+P_{\lambda}(X_{2}=+1) =1$
 is a relation which holds for any $\lambda\in \Lambda$, while the third equality descends
 from Eq.~(\ref{eq4.54}).
Finally, since $P_{\sigma}(Y_{1}=+1,X_{2}=-1)$ equals $\varepsilon$ due to Eq.~(\ref{eq4.52}), we
 obtain the desired relation between the indicated subsets $C$ and $B\cap C$ of $\Lambda$, that is
\begin{equation}
\label{eq4.61} \mu [C] - \mu [B\cap C] = 2\varepsilon\:.
\end{equation}
Similar arguments can be used with the other probability distributions of Eqs.~(\ref{eq4.51})-(\ref{eq4.56})
 so as to obtain constraints which the measures of appropriate subsets of $\Lambda$ have to satisfy if
 a local realistic model for $\sigma$ exists, that is
\begin{eqnarray}
 \label{eq4.71}
 \mu[A \cap B] &= & \varepsilon,\\
 \label{eq4.72}
 \mu[C] - \mu[ B\cap C] & = & 2\varepsilon,\\
\label{eq4.73}
 \mu[D]- \mu[A\cap D] & = & 2\varepsilon,\\
\label{eq4.74}
 \mu[ C\cap D] & = & a+\varepsilon.
\end{eqnarray}
These equations are similar to Eqs.~(\ref{eq1.91})-(\ref{eq1.94}), apart from a multiplicative
 factor which appears in Eqs.~(\ref{eq4.72}) and~(\ref{eq4.73}) and which is related to the fact the observables
 $X_{i}$ and $Y_{j}$ we are considering, possess now three different eigenvalues, rather than two as before.
Due to this similarity, we can follow the set-theoretic manipulations we used in Ref.~\cite{gm2} and
 conclude that if a local realistic model exists for the mixed states of Eq.~(\ref{eq4.2}), then the following
 inequality, involving the parameters $\varepsilon$ and $a$, has to hold
\begin{equation}
\label{eq4.8}
 4\varepsilon - a\geq 0.
\end{equation}
Taking into account the definition of $\varepsilon$ and $a$ given in Eqs.~(\ref{eq4.51}) and~(\ref{eq4.56})
 with respect to $d_{1},d_{2},p_{1},p_{2}$, and $p$, the following theorem follows:\\

{\bf Theorem III}. Consider the normalized entangled vector $\vert \phi \rangle = \sum_{i\geq 1}p_{i}\vert
 i-1\rangle
 \otimes \vert i-1 \rangle$ belonging to ${\mathbb{C}}^{d_{1}}\otimes {\mathbb{C}}^{d_{2}}$,
  with different (strictly positive) weights $p_{1}\neq p_{2}$, and
 the one-parameter class of mixed statistical operators $\sigma = p\vert \phi\rangle \langle \phi
 \vert + \frac{1-p}{d_{1}{d_2}}I_{d_{1}}\otimes I_{d_{2}}$, where $p\in [0,1]$.
If there exists a local and deterministic hidden variable model for $\sigma$ then
 $ p \in [0, \frac{1}{1+ \frac{d_{1}d_{2}p_{1}^{2}p_{2}^{2}(p_{1}-p_{2})^2}{4(p_{1}^{2}+p_{2}^{2}-p_{1}p_{2})^{2}}}]$.\\

Once again, this theorem allows us to conclude that the one-parameter class of statistical
 operators $\sigma$ of Eq.~(\ref{eq4.2}) with values of $p$ such that
\begin{equation}
\label{eq5.0} p\in (\frac{1}{1+
\frac{d_{1}d_{2}p_{1}^{2}p_{2}^{2}(p_{1}-p_{2})^2}{4(p_{1}^{2}+p_{2}^{2}-p_{1}p_{2})^{2}}},1],
\end{equation}
cannot be described by any local realistic model. This naturally implies that such states are
 also not separable.

In this general scenario, where arbitrary Hardy states belonging to ${\mathbb{C}}^{d_{1}}\otimes
 {\mathbb{C}}^{d_{2}}$ are mixed with a completely chaotic noise, a (necessary)
 condition, like that of Ref.~\cite{horo}, on the values of $p$ for the existence of a local realistic model,
 is not known yet.
As a consequence, contrary to what happened when dealing with the two-qubit case where
 the result proven in Ref.~\cite{horo} could have been applied, it is not possible to compare the
 strength of our result with some (alternative) criterion based on the violation of a Bell-like inequality,
 because the latter has not been discovered yet.
Thus, we have succeeded to discover whole classes of mixed states, like those of Eq.~(\ref{eq4.2})
 where $\vert \phi\rangle$ is any Hardy state whatsoever, such that (i) no local model may exist for them,
 (ii) they are proven to be nonseparable, without having considered any criterion based on Bell inequalities.

Moreover, in this paper we have considerably enlarged, with respect to the results obtained in Ref.~\cite{gm2},
 the interval of values for the parameter $p$ such that the corresponding mixed states do not admit any
 local hidden variable model.
In fact, in Ref.~\cite{gm2} a general argument to establish the nonlocal features of certain
 classes of mixed states, involving set-theoretic techniques similar to those we used in this paper,
 has been exhibited.
In that paper we have proved that, given the Hardy state $\vert \phi\rangle$ of Eq.~(\ref{eq4.1}) and
 an arbitrary mixed state $\sigma$ whose trace distance $D(\sigma, \vert \phi\rangle\langle \phi\vert )=
 \frac{1}{2}\textrm{Tr}\vert (\sigma - \vert \phi\rangle \langle \phi\vert)\vert$ from the pure state
 $\vert \phi\rangle\langle \phi\vert$ we have  denoted as $\eta$, if
\begin{equation}
 \label{eq6.0}
 0 \leq \eta <\frac{p_{1}^{2}p_{2}^{2}(p_{1}-p_{2})^2}{6(p_{1}^{2}+p_{2}^{2}-p_{1}p_{2})^{2}}\:,
 \end{equation}
then no local realistic model exists for $\sigma$. The idea of Ref.~\cite{gm2} was that in a definite
 neighborhood - with respect to the topology induced by the consideration of the trace distance - of a Hardy
 state there exist uncountable many mixed states which exhibit nonlocal features, and we succeeded in
 determining the size of that neighborhood.
Unfortunately, the result of Eq.~(\ref{eq6.0}) was not an optimal one because it did not rely directly on the
 specific form of the mixed state $\sigma$: in fact, in order to be completely general, in Ref.~\cite{gm2} we
 resorted to majorizations, based on the consideration of the trace distance, for the probability distributions
 of Eqs.~(\ref{eq4.51})-(\ref{eq4.56}).
On the contrary, the method we have presented in this paper makes explicit use of the expression of $\sigma$ of
 Eq.~(\ref{eq4.2}), which represents a Hardy state corrupted by a white noise,
 to calculate exactly the relevant probability distributions.
As a consequence, it provides us with a larger range of values of the parameter $p$
 for which the corresponding statistical operators do not admit a local realistic model, just because the proof
 relies on the specific properties of $\sigma$.
In fact, given the state of Eq.~(\ref{eq4.2}), we may easily evaluate
\begin{equation}
\label{eq7.0}
 D(\sigma, \vert \phi\rangle\langle \phi\vert ) = (1-p)\frac{d_{1}d_{2}-1}{d_{1}d_{2}} \equiv \eta
\end{equation}
and, due to the result of Eq.~(\ref{eq6.0}), this implies that whenever $p\in
(1-\frac{d_{1}d_{2}}{6(d_{1}d_{2}-1)}
\frac{p_{1}^{2}p_{2}^{2}(p_{1}-p_{2})^2}{(p_{1}^{2}+p_{2}^{2}-p_{1}p_{2})^{2}},1]$ the associated
 statistical operators cannot be mimicked by local deterministic hidden variable models.
The fact that the criterion presented in this paper is more powerful than the one of Ref.~\cite{gm2}
 is apparent since the inequality
\begin{equation}
\label{eq7.1} \frac{1}{1+
\frac{d_{1}d_{2}p_{1}^{2}p_{2}^{2}(p_{1}-p_{2})^2}{4(p_{1}^{2}+p_{2}^{2}-p_{1}p_{2})^{2}}}<
 1-\frac{d_{1}d_{2}}{6(d_{1}d_{2}-1)}
\frac{p_{1}^{2}p_{2}^{2}(p_{1}-p_{2})^2}{(p_{1}^{2}+p_{2}^{2}-p_{1}p_{2})^{2}}
\end{equation}
holds for single-particle Hilbert spaces of dimension $d_{i}$ greater or equal to $2$ and for any value of
 $p_{1},p_{2}\in [0,1]$ and its validity may be established by very simple analytical calculations.
Yet, to better appreciate the improvement with respect to the result of Ref.~\cite{gm2}, we have plotted the
 values of $1-p=1-\frac{1}{1+\frac{d_{1}d_{2}p_{1}^{2}p_{2}^{2}(p_{1}-p_{2})^2}{4(p_{1}^{2}+p_{2}^{2}-p_{1}p_{2})^{2}}}$
 (the upper curves) and of $1-p=\frac{d_{1}d_{2}}{6(d_{1}d_{2}-1)}
 \frac{p_{1}^{2}p_{2}^{2}(p_{1}-p_{2})^2}{(p_{1}^{2}+p_{2}^{2}-p_{1}p_{2})^{2}}$ (the lower curves)
 providing evidence of nonlocality, versus the parameter $p_1$ for both criteria and for arbitrarily
 chosen values of $d_{1}d_{2}$ and $p_2$:

\begin{figure}[th]
 \centerline{\epsfig{file=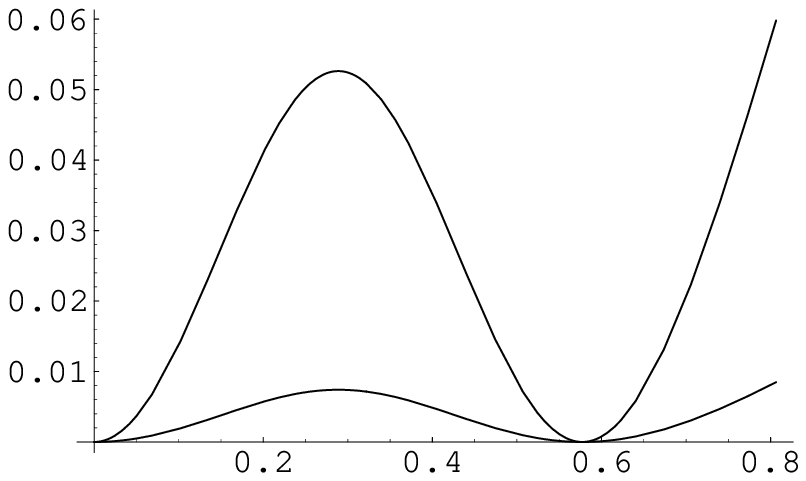, width=5cm} \hspace{0.2cm}
 \epsfig{file=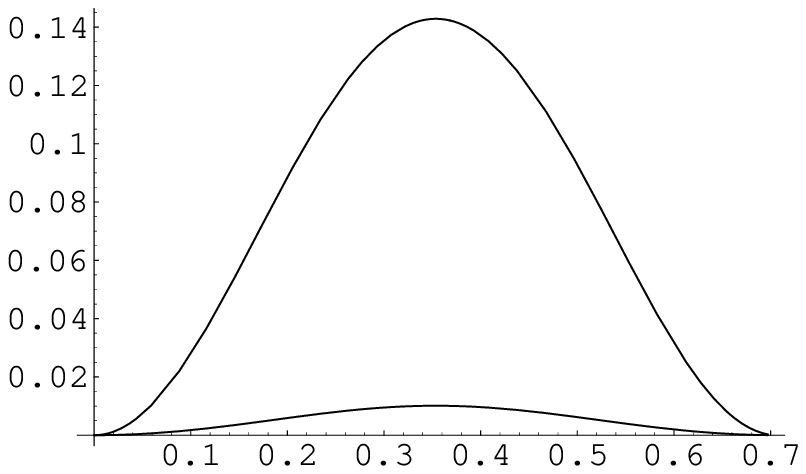, width=5cm}\vspace*{8pt}}
 \caption{\label{fig1} Values of $1-p$ versus $p_{1}$, for $d_{1}d_{2}=6$ and $p_2=1/\sqrt{3}$
 (left figure) and for $d_{1}d_{2}=12$ and $p_{2}=1/\sqrt{2}$, with the constraint that $p_{1}^{2}+p_{2}^{2}\leq
 1$.}
\end{figure}

Since the greater are the values of $1-p$ (plotted in the vertical axis) the larger is the set of mixed states
 exhibiting nonlocal features, the class of mixed states not admitting local descriptions discovered in this paper
 (the upper curves in Fig.1) is appreciably bigger than the one obtained through the the trace-distance method of
 Ref.~\cite{gm2} (the lower curves in Fig.1).
Similar results can be obtained for any choice of $d_1d_2>2$ and for any $p_2$. The considerable improvement we
 have achieved here had to be expected because in Ref.~\cite{gm2} we have not restricted in any way the nature of
 the noise.


\section{Conclusions}

We have shown how a refinement and a generalization of the original Hardy proof of nonlocality
 without inequalities can be used to deny the existence of local and deterministic hidden
 variable models for some one-parameter classes of mixed statistical operators.
Such classes contain convex mixtures of pure Hardy states with a completely chaotic
 noise or a colored noise, and they represent typical mixed states which are considered in the experimental
 realizations of nonlocality tests.
We have explicitly exhibited precise ranges of values of the parameter measuring the amount
 of noise affecting a nonoptimal preparation of a (pure) Hardy state for which nonlocal
 effects can be still experimentally revealed and, in some cases, we have bettered the results
 obtained in Ref.~\cite{gm2}.


\section{Acknowledgments}
Work supported in part by Istituto Nazionale di
  Fisica Nucleare, Sezione di Trieste, Italy.


\end{document}